\UseRawInputEncoding
\pdfoutput=1
\documentclass[twocolumn,aps,prl,superscriptaddress,notitlepage,floatfix,longbibliography]{revtex4-2}

\usepackage{graphicx}
\usepackage{epsfig}
\usepackage{float}
\usepackage{dcolumn}
\usepackage{bm}
\usepackage{amsmath}
\usepackage{float}
\usepackage{amssymb}
\usepackage{lipsum}
\usepackage{titletoc}
\usepackage{makecell}

\usepackage{array}
\usepackage{color}
\usepackage{bbding}

\begin{document}

\title{Anomalous open orbits in Hofstadter spectrum of Chern insulator}%

\author{Haijiao Ji}
\affiliation{Center for Advanced Quantum Studies, Department of Physics, Beijing Normal University, Beijing 100875, China}
\affiliation{Harbin Institute of Technology, Shenzhen, 518055, P. R. China}
\author{Noah F. Q. Yuan}
\affiliation{Harbin Institute of Technology, Shenzhen, 518055, P. R. China}
\author{Hua Jiang}
\affiliation{School of Physical Science and Technology, Soochow University, Suzhou 215006, China}
\affiliation{Institute for Advanced Study, Soochow University, Suzhou 215006, China}
\author{Haiwen Liu}
\email[Haiwen Liu:]{haiwen.liu@bnu.edu.cn}
\affiliation{Center for Advanced Quantum Studies, Department of Physics, Beijing Normal University, Beijing 100875, China}
\author{X. C. Xie}
\affiliation{International Center for Quantum Materials, School of Physics, Peking University, Beijing 100871, China}
\affiliation{CAS Center for Excellence in Topological Quantum Computation}

\begin{abstract}
The nontrivial band topology can influence the Hofstadter spectrum. We investigate the Hofstadter spectrum for various models of Chern insulators  under a rational flux $\frac{\phi_{0}}{q}$, here $\phi_{0}=\frac{h}{e}$ and $q$ being an integer. We find two major features. First, the number of splitting subbands is $|q-C|$ with Chern number $C$. Second, the anomalous open-orbit subbands with Chern numbers $q-1$ and $-q-1$ emerge, which are beyond the parameter window $(-q/2,q/2)$ of the Diophantine equation studied by Thouless-Kohmoto-Nightingale-den Nijs [Phys. Rev. Lett. \textbf{49}, 405 (1982)]. These two findings are explained by semiclassical dynamics. We propose that the number of splitting subbands can be utilized to determine Chern number in cold atom systems, and the open-orbit subbands can provide routes to study exotic features beyond the Landau level physics.
\end{abstract}
\maketitle

\textit{Introduction.} --- The simultaneous presence of lattice and magnetic length gives rise to the celebrated quantum fractal--the Hofstadter spectrum\cite{hofstadter1976energy,satija2016butterfly}, which is beyond the scope of band theory. Recently, experimental realization of Hofstadter spectrum in twisted bilayer graphene\cite{dean2013hofstadter,hunt2013massive,ponomarenko2013cloning,andrei2020graphene,carr2017twistronics} and cold atom systems\cite{roati2008anderson,deissler2010delocalization,aidelsburger2013realization,wang2013proposal,schreiber2015observation,bordia2017probing,zhou2021thermalization} have generated extensive interests to investigate this exotic spectrum in synthetic Chern insulators\cite{wu2016realization,pierce2021unconventional,lu2021multiple,saito2021hofstadter,finney2022unusual,hauke2022quantum}.
Moreover, for magic angle twisted bilayer graphene, recent studies have identified a series of robust correlated Chern insulators and transport measurements suggest the possibility of a new van Hove singularity and fractional Chern insulator at certain condition\cite{hauke2022quantum,wu2021chern}. 
The interpretation of these rich physics need the investigation of Hofstadter spectrum of Chern insulator, which remains poorly understood theoretically.

\begin{figure}
\includegraphics[width=0.40\textwidth]{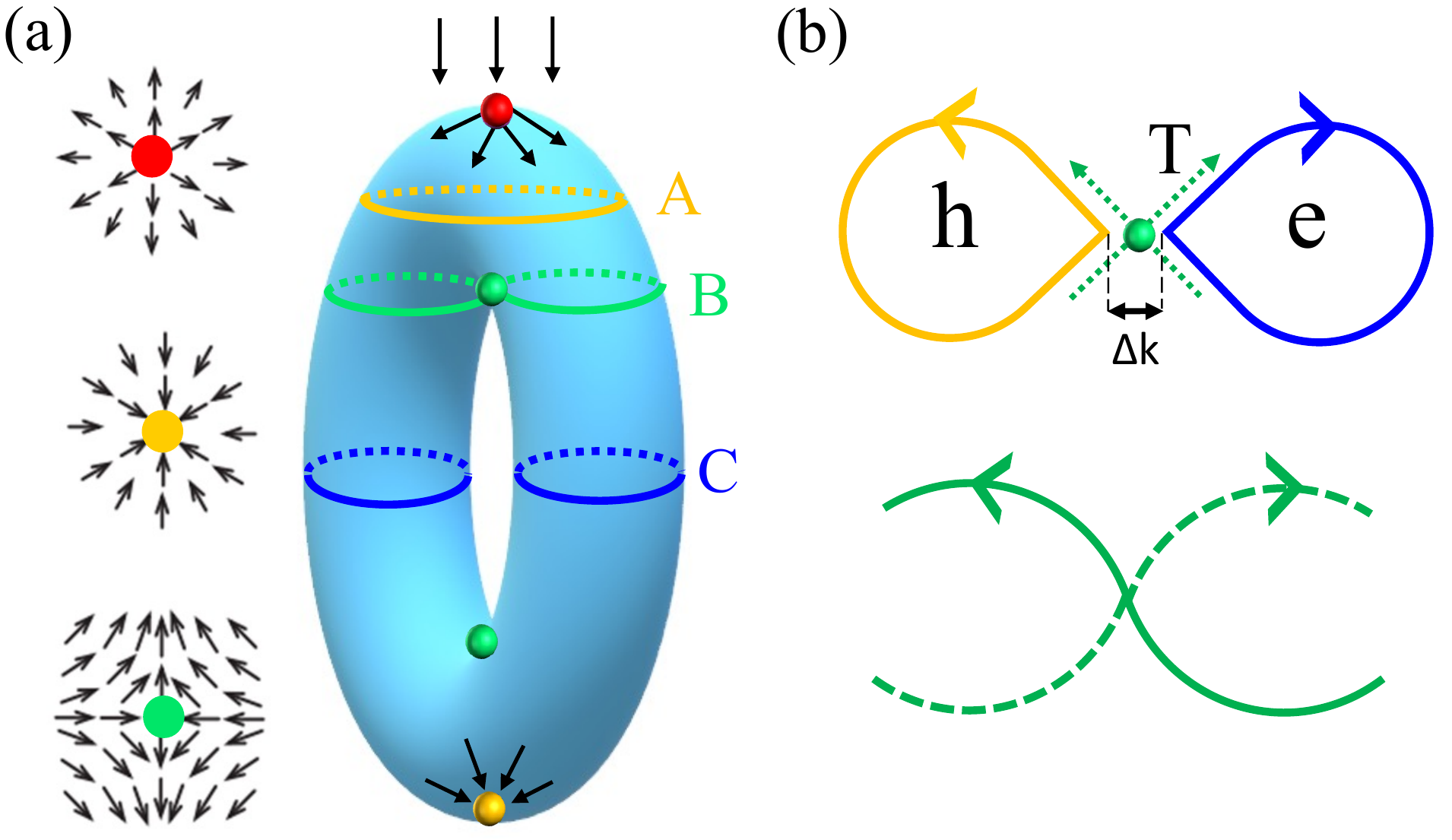}
 \caption{(a) Velocity field near maximum point(red), minimum point(yellow), saddle point(green). Closed orbit $A$ is contractible, and open orbits $B$ and $C$ are non-contractible. Orbit $B$ is a self intersecting orbit. (b) The quantum tunneling $T$ near saddle point $(\Delta k\rightarrow 0)$ induce the anomalous open orbits with opposite directions of propagation, reminiscent of the magnetic breakdown effect (the details are shown in Fig. \ref{fig4}). 
  }\label{fig1}
\end{figure}

For conventional Hofstadter spectrum, there exists two theoretical methods to describe. One is the algebraic Diophantine equation proposed by Wannier\cite{wannier1978result}, $n/n_{0}=t\phi/\phi_{0}+s$, which describe the relationship between carries densities $n/n_0$ and magnetic flux $\phi$, where $t$ and $s$ are two integers and $\phi_{0}$ is the flux quanta. P. Streda demonstrated $s$ is filling factor of Bloch band\cite{streda1982theory}, and Thouless-Kohmoto-Nightingale-den Nijs (TKNN) have demonstrated $t$ is the topological invariant corresponding to the quantized Hall conductance by relation $\sigma_{xy}=t\frac{e^2}{h}$\cite{thouless1982quantized}. Landau fan diagram obtained in transport experiment is the direct display of Diophantine equation.
The other is semiclassical depiction and the investigation of closed and open orbit is a key\cite{chang1995,chang1996berry}. Considering the Brillouin zone as a torus in FIG. \ref{fig1}(a), the closed orbit (like orbit A) locates around the maximum or minimum, and can shrink into a point. The open orbits emerge around the saddle-points of energy spectrum\cite{roth1966semiclassical}, and penetrate throughout the torus (orbit B or C). Under a magnetic field, the quantum tunneling between electron pocket and hole pocket around the saddle point can diminish two closed orbits and gives birth to two open orbits, reminiscent of the magnetic breakdown process \cite{kaganov1983coherent,o2016magnetic,linnartz2022fermi,alexandradinata2018semiclassical}, as shown in FIG. \ref{fig1}(b). In reality, the semiclassical orbits evolve into hyperorbital subbands. The closed-orbit subbands can be adiabatically connected with Landau levels, whereas the open-orbit subbands are distinctive from Landau levels with Chern number breaking the parameter constraint in the TKNN consideration\cite{chang1995,chang1996berry}. 
The Hofstadter spectrum in Chern insulators and the associated open-orbit subbands may go beyond the TKNN constraint, and theoretical investigations on these issues can enrich the understanding of this exotic Hofstadter spectrum.  

\begin{figure}
\centering
\includegraphics[width=0.47\textwidth]{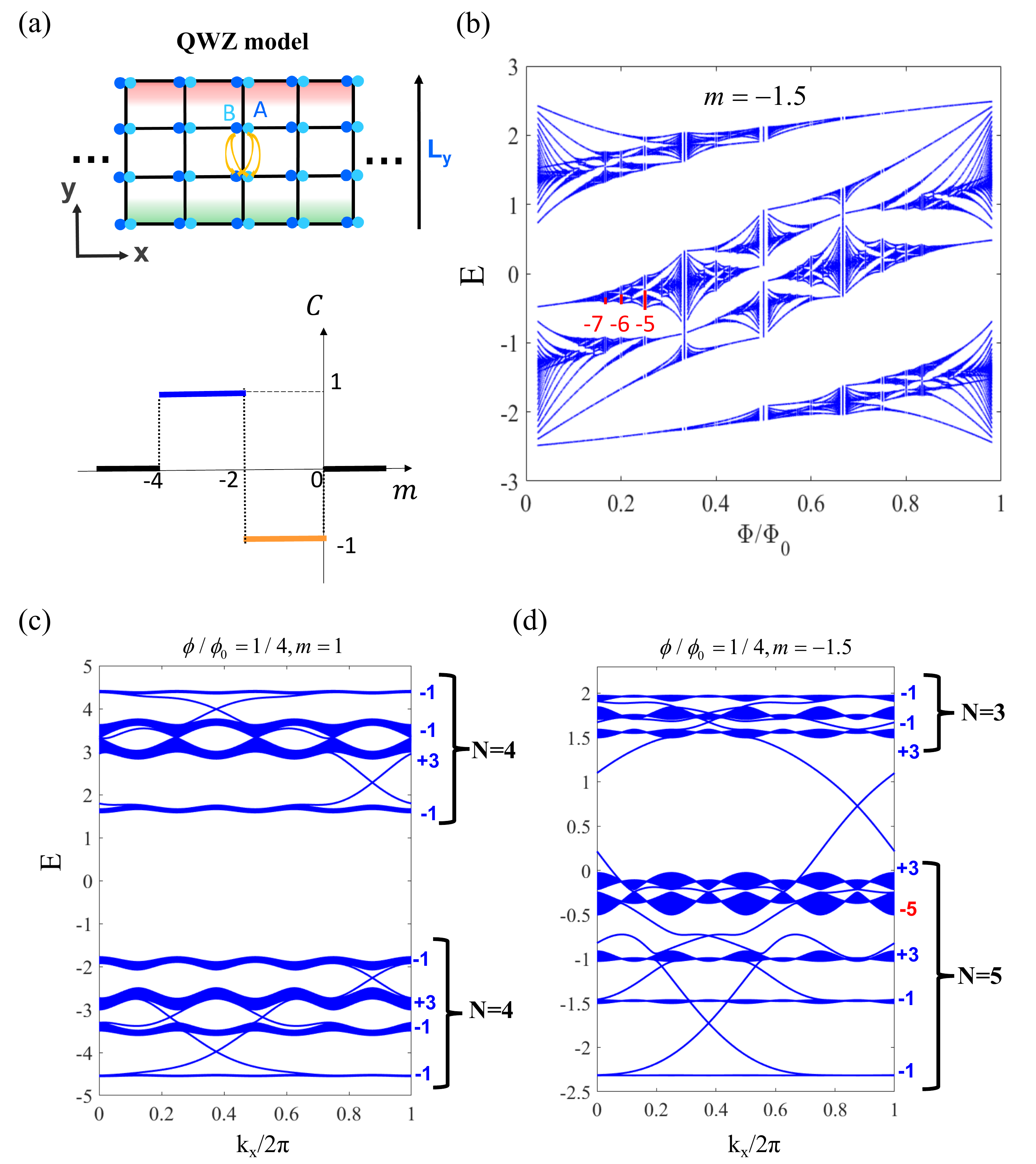}
 \caption{(a) The schematic and phase diagram of the QWZ model, $A$ and $B$ denote two atomic orbits. Here, we consider a ribbon geometry with open boundary along $y$ direction with width $L_y$. (b) The Hofstadter spectrum of QWZ model at parameter $m=-1.5$. Red line label the exotic open-orbit subbands with Chern number $-q-1$. (c) The energy spectrum for QWZ model with trivial topology band at $m=1$ under flux $\phi/\phi_0=1/4$. The blue numbers label the Chern numbers of the splitting subbands. The largest energy gap is the 4-th gap which separate whole spectrum into two parts.
 (d) The spectrum for non-trivial topology band at $m=-1.5$ under flux $\phi/\phi_0=1/4$. The largest energy gap is the 5-th gap. The top three subbands originate from the higher energy parent band with Chern number $C=+1$, while the other five subbands are from the lower energy parent band with Chern number $C=-1$. Comparing to (c), anomalous open-orbit subbands with $C=-5$ appear in (d). 
  }\label{fig2}
\end{figure}

Here we investigate the Hofstadter spectrum of Chern insulator under a rational magnetic flux $\phi/\phi_{0}=1/q$ for the Qi-Wu-Zhang(QWZ) model. We find out the number of splitting subbands $|q-C|$ is determined by Chern number $C$ of parent band, and anomalous open orbits with Chern number $q-1$ and $-q-1$ emerge around the saddle points of the parent band. These results are also verified for Haldane model and higher Chern number Chern insulator model, and the realization of open-orbit subbands are experimentally feasible in twisted bilayer graphene and cold atom systems.

\textit{The Hofstadter spectrum of QWZ model.} -- Firstly, we consider the QWZ model, which possess nonzero Chern number in certain parameter regime, and is described by two-sublattice tight-binding Hamiltonian\cite{qi2006topological}
\begin{equation}
{H} =\sum_{i} c_{{i}}^{\dagger}Mc_{{i}}
-\sum_{{i}} (c_{i}^{\dagger} t_{x}c_{i+x} +c_{i}^{\dagger}t_{y} c_{i+y}+{ h.c. })
 \label{eq1}
\end{equation}
where $c_{i}$ is two-component spinor, the onsite mass term is $M=(m+2)\sigma_z$, nearest neighbour hopping terms are $t_{x,y}=\frac{1}{2}(\sigma_{z}+i \sigma_{x,y})$, and Pauli matrices $\sigma_{x,y,z}$ act on the sublattice space. The Fourier transform of Eq. (\ref{eq1}) reads $\mathcal{H}(\vec{k}) =\bm d(\vec{k}) \cdot\bm\sigma$ with
$d_{x}(\vec{k})=\sin k_{x}$,
$d_{y}(\vec{k}) =\sin k_{y}$,
$d_{z}(\vec{k}) = (m+2)-(\cos k_{x}+\cos k_{y})$.
Two particle-hole symmetric bands $\pm|\bm d|$ are hence found for the QWZ model.
The Chern number of the lower band $-|\bm d|$ can be obtained via mapping from the toric Brillouin zone (BZ) to the sphere $C=\frac{1}{4 \pi} \int_{\rm BZ} d k_{x} d k_{y} \frac{\partial \hat{{\bm d}}}{\partial k_{x}} \times \frac{\partial \hat{{\bm d}}}{\partial k_{y}} \cdot \hat{{\bm d}}$\cite{thouless1982quantized}, which is only determined by the value of $m$, as depicted in FIG. \ref{fig2}(a).
Three phases can be found, corresponding to $C=0,\pm 1$.
At the topological phase transition points $C$ is ill-defined, as the upper and lower bands touch at Dirac points.

Under an out-of-plane magnetic filed $B\hat{\bm z}$, the Peierls phase factor $\exp\left({i \frac{e}{\hbar} \int_{i}^{j} \bm{A} \cdot d \bm{l}}\right)=e^{i 2 \pi \frac{\phi}{\phi_{0}}}$ needs to be introduced in the hopping term, which induces the non-commutativity of translation operators along $x$ and $y$ direction. 
However, if the magnetic flux of unit cell satisfies $\phi/\phi_{0}=p/q$ (where $p$ and $q$ are co-prime integers), one can enlarge the unit cell by $q$ times (also known as the magnetic unit cell) to restore the commutativity of lattice translation symmetry. It implies that a Bloch band splits into $q$ magnetic Bloch subbands under magnetic field. 
Detailed calculations on magnetic Bloch subbands are provided in Section I of the Supplementary Material (SM). 

In order to study the Chern numbers of subbands in the Hofstadter spectrum, we consider a ribbon geometry with open boundary along $y$ direction and use the gauge $A=(-By,0)$.
The energy spectrum of this model under flux $p/q=1/4$ As shown in FIG. \ref{fig2}(c, d), two parent bands of QWZ ribbon model split into $2q$ subbands under and edge states appear inside the subband gaps. The Hall conductance inside the gaps of the sub-structure can be calculated by Kubo formula\cite{thouless1982quantized}. The Chern number of splitting subbands can be derived from Hall conductance according to bulk-edge correspondence $\sigma_{xy}^{bulk}=-\frac{e^2}{h}C^{edge}$ \cite{rudner2013anomalous,graf2013bulk,hatsugai2009bulk} and the sum rule\cite{simon1983holonomy,avron1985stability}. The detailed explanations are given in Section II of the SM.

\textit{The Chern numbers sequence of subbands in Hofstadter spectrum.} ---
In FIG. \ref{fig2}(c) and (d), we investigate the trivial and nontrivial parent bands at $m=1\ (C=0)$ and $m=-1.5\ (C=-1)$ respectively, where the magnetic flux is ${p}/{q}={1}/{4}$, and the largest band gap near zero energy separates upper and lower parent bands. For our choice of $q=4$, we find the lower parent band split into $N=4,5$ subbands when $C=0,-1$ respectively.
Moreover, in the trivial case $C=0$ of FIG. \ref{fig2}(c), two of subbands carry Chern number $+3$ and the rest subbands carry Chern number $-1$. 
In the topological case $C=-1$ of FIG. \ref{fig2}(d), subbands with new Chern number $-5$ appear. Similar anomalous subbands with Chern numbers $-7,-6,-5$ is verified at various $q=6,5,4$, marked by red lines in the Hofstadter spectrum as shown in FIG. \ref{fig2} (b).
The overall sequences at different parameters are summarized in Table. IV in the SM.

From the numerical results of Chern numbers sequence, we extract two major findings. First, the number of splitting subbands is $|q-C|$, which is related to the Chern number $C$ of parent band. Second, the anomalous open-orbit subbands with non-unity Chern number $q-1$ and $-q-1$ emerge, which are beyond Landau level picture and the parameter window $(-q/2,q/2)$ of the Diophantine equation considered by Thouless-Kohmoto-Nightingale-den Nijs\cite{thouless1982quantized}. Moreover, to verify the generality and robustness of these findings, we also consider the Hofstadter spectrum of Haldane model and higher Chern number Chern insulator (High-C) model, which can be realized experimentally in graphene system and optical cold atom system. These specific models and results are provided in Section I and III in the SM. And above two findings do hold in all of the three models (QWZ, Haldane, High-C).
In the following sections, we offer semiclassical understandings for these two exotic properties.

\textit{Number of splitting subbands.} ---
For $C=0$ QWZ model with rational flux $p/q$, the topological number of subbands are the direct combination of two simple square lattice, which satisfy the algebraic Diophantine equation within the TKNN parameter regime\cite{thouless1982quantized}
\begin{equation}
    r=pt+qs,\quad t\in(-q/2,q/2),
\end{equation}
where $t$ is quantum number of Hall conductivity in the $r$th gap and $s$ is another quantum number about the filling of Bloch electron. However, the results of $C\neq0$ QWZ model are inconsistent with the solutions of Diophantine equation in the regime $\sigma \in (-q/2,q/2)$ as described in the SM. 
We turn to the semiclassical analysis of the subbands based on the semiclassical dynamics of hyperorbits\cite{chang1995,chang1996berry}.

\begin{figure}[htpb]
\includegraphics[width=0.46\textwidth]{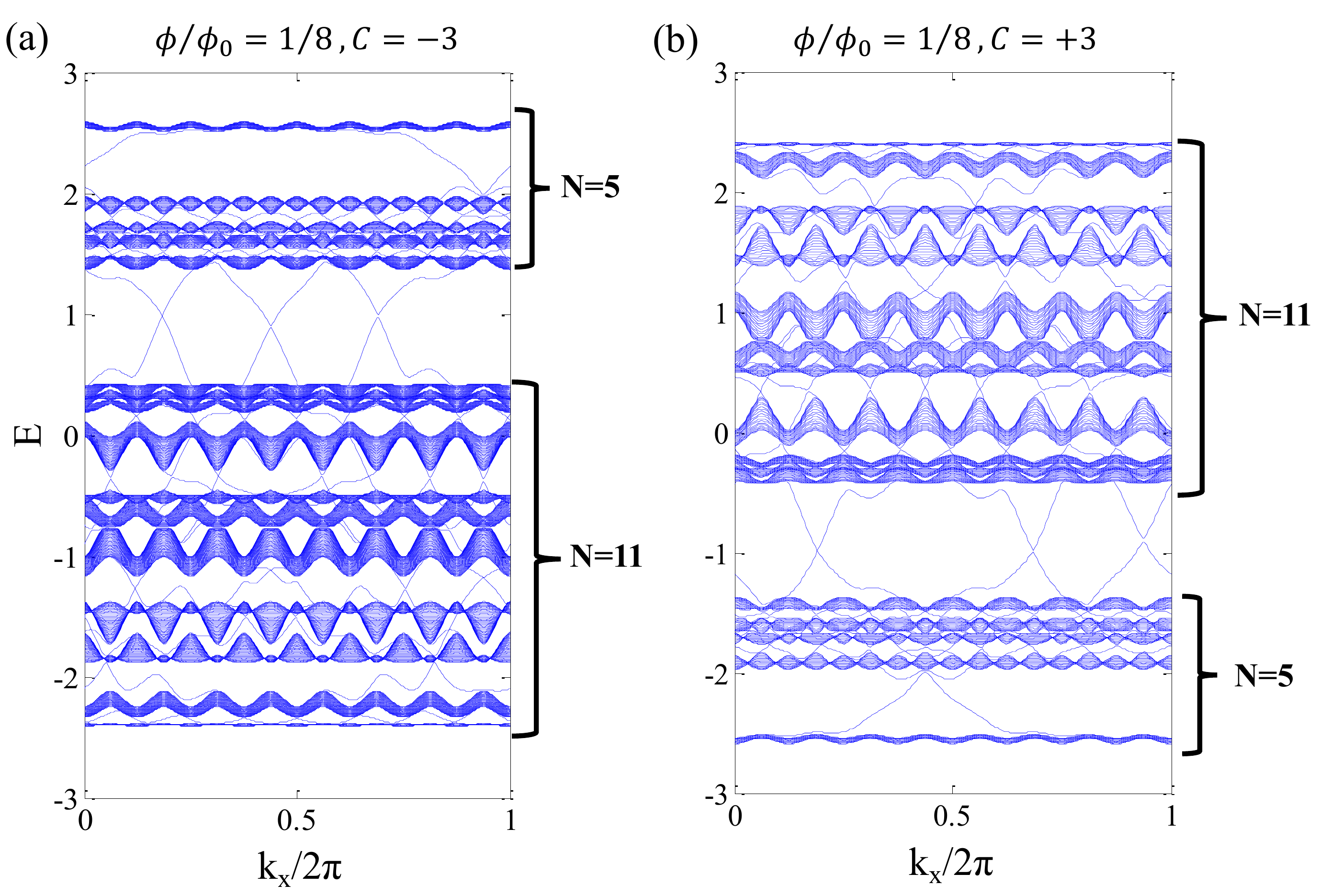}
 \caption{(a) The energy spectrum of higher Chern insulator model at $C=-3$ under flux $\phi/\phi_0=1/8$. The lower parent band split into $N=11$ subbands. (b) same as (a) but at $C=+3$. The lower parent band split into $N=5$ subbands.
  }\label{fig3}
\end{figure}

In the semiclassical consideration, the external $\bm{B}$ can be divided as $\bm{B}=\bm{B}_{0}+\delta \bm{B}$, in which $\bm{B}_{0}$ gives rise to magnetic Bloch bands and $\delta \bm{B}$ represents a perturbation to the cyclotron motion of magnetic Bloch electrons. To differentiate them from the usual orbits of Bloch electrons, Chang and Niu\cite{chang1995,chang1996berry} introduced the concept of “hyperorbit" whose equation of motion in $k$ space is $\hbar \dot{\bm{k}}=-e Z_{\delta B}(\bm{k}) \frac{\partial \varepsilon (\bm{k})}{\hbar \partial \bm{k}} \times \delta \bm{B}$, where $Z_{\delta B}=(1+\Omega(\bm{k})\delta Be/\hbar)^{-1}$ is a curvature correction factor and $\Omega(\bm{k})$ is the “Berry curvature”.
The quantization rule for hyperorbit $\mathcal{O}$ in the magnetic Bloch band reads $\frac{1}{2} \oint_{\mathcal{O}}(\bm{k} \times d \bm{k}) \cdot \hat{\bm z}=\left[(2m+1)\pi-{\Gamma\left(\mathcal{O}\right)}\right] \frac{e \delta B}{\hbar}$, and $\Gamma\left(\mathcal{O}\right)=\oint_{\mathcal{O}} \mathcal{A} \cdot d \mathbf{k}$ is the Berry phase for $\mathcal{O}$. This quantization rule determines the number of subbands in the Hofstadter spectrum. For $B_{0}S = {p_{0}}/{q_{0}}$ with the unit cell area $S$, the number of quantized hyperorbits in first magnetic Brillouin zone is
\begin{equation}
    N=\frac{\left|1 /\left(q_{0} \delta \phi\right)-C\right|}{q_{0}} =|q-C|
    \label{eq19}
\end{equation}
where $\delta \phi=\delta {B}S=1/{q}$, $C$ is the Chern number of parent band and $p_0/q_{0}=0/1$ is the magnetic flux of $B_{0}=0$. 
Eq. (\ref{eq19}) also be applied to the high Chern number model in FIG. S\ref{fig2} in SM, where the parent bands with $C=-3, +3$ split into $N=11, 5$ subbands respectively at flux $p/q=1/{8}$.
It is worth mentioning that the parent bands at zero field possess nontrivial topology in this work, which are different from the models shown in Refs.\cite{chang1995,chang1996berry}. Thus, our work open routes to use the number of splitting subbands as a criterion to demarcate the topological properties of constructed bands in cold atom system\cite{aidelsburger2013realization,wu2016realization,hauke2022quantum}.

\begin{figure*}
\includegraphics[width=0.98\textwidth]{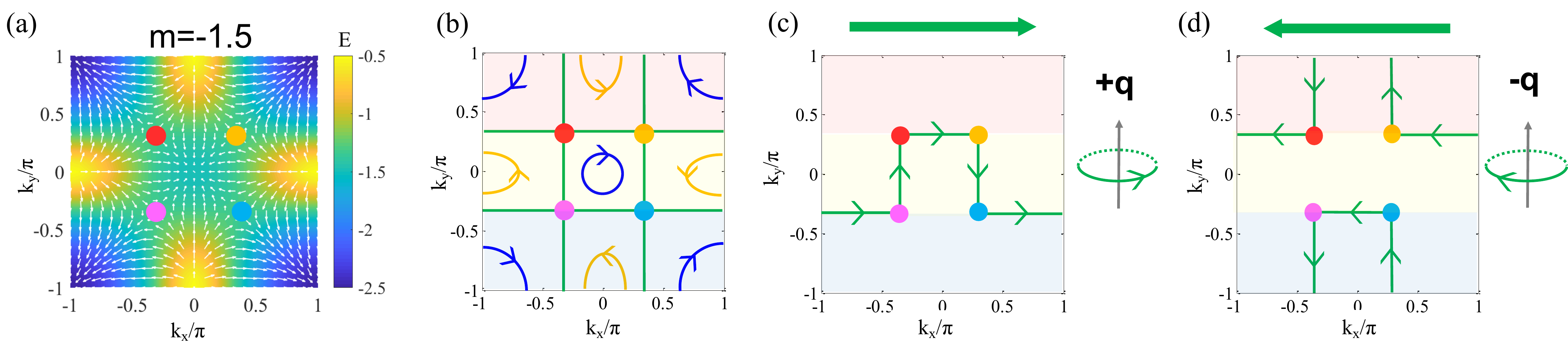}
 \caption{(a) The energy spectrum of lower band in QWZ model at $m=-1.5$. White arrows label the vector fields of spectra. Red, orange, purple and blue points label the saddle points of energy spectra. (b) The constant energy contours in the Brillouin zone of parent Bloch band at the same parameter with (a). Here, yellow and blue lines label the closed orbits in the electron pockets and hole pockets, and green lines label the open orbits. For the case of flux $p/q=1/3$, the Brillouin zone fold into three smaller magnetic Brillouin zone filled by blue, yellow and red colors. (c,d) The trajectory of open orbits in (b) can be separated into two opposite orbits with Chern number $\pm q$, respectively. Here take $k_x$ and $-k_x$ direction as examples. These schematics demonstrate the open-orbit contribution to Chern number (first term in Eq. \ref{eq4}) in addition to the conventional Streda formula (second term in Eq. \ref{eq4}).
 \label{fig4}}
\end{figure*}

\textit{Anomalous open-orbit subbands.} --- In periodic systems, the hyperorbits will be classified as “closed" or “open" ones. All of the closed-orbit subbands at fixed $1/q$ have the same Chern number $-1$, which can be obtained from $\sigma^{close}={e \rho}/{\delta B}=1/({q\delta \phi})$\cite{chang1996berry}. In order to satisfy the sum rule $\sigma_{\text {parent}}=\sum \sigma_{\text {daughter}}$, Chang and Niu demonstrate that there is only one open-orbit subband with non-unity chern number $q-1$ for every parent band in a square lattice model where only nearest neighbor hopping are considered.


However, as shown in FIG. \ref{fig2}(d), we find four open-orbit subbands in the Hofstadter spectrum of QWZ model. Three of them carry Chern number $q-1=+3$ as described by Chang and Niu, while one anomalous open-orbit subband with Chern number $-q-1=-5$ (marked red) emerges, which is beyond the previous semiclassical description\cite{chang1996berry}. During the topological transition of parent bands from FIG. \ref{fig2}(c) to (d), one open-orbit subband with Chern number $-1$ in the upper parent band merge to the lower parent band, producing two open-orbit subbands with Chern numbers $-q-1$ and $+q-1$ respectively in the lower parent band.
In Tables IV,V and VI in the SM, we summarize the topological transitions of open-orbit subbands and the emergence of anomalous open-orbit subbands with $-q-1$ in the Hofstadter spectrum of QWZ and Haldane models. Above results imply the semiclassical picture proposed by Chang and Niu\cite{chang1996berry} requires generalization to account for the behaviour of the anomalous open orbits.

From the semiclassical point of view, near the saddle point of energy spectrum, some closed orbits from the electron pockets and some closed orbits from the hole pockets can hybridize due to quantum tunneling, and hence evolve into open orbits. Such a process is similar to the phenomena of magnetic breakdown, 
such as the magnetic breakdown process between electron and hole bands in type-II Weyl semimetal WTe$_{2}$\cite{linnartz2022fermi,alexandradinata2018semiclassical}. 
Here we consider a similar tunneling process within a single band around the saddle point, and focus on the Chern numbers of the anomalous open-orbit subbands. 

From the semiclassical equation of motion $\hbar{\dot{\bm k}}=-e{\bm E}-e{\dot{\bm r}}\times\delta {\bm B}$, 
one can obtain the current density of subbands\cite{chang1996berry}
\begin{equation}
{\bm J}=-\int d^2{\bm k} \frac{\hbar}{\delta B}{\bm{\dot k}}\times {\hat{\bm z}}-e
\int d^2{\bm k} \frac{{\bm E}\times{\hat{\bm z}}}{\delta B}.
\label{eq4}
\end{equation}
The second term is normal drift term of wave packet center, which is identical to the result of the Streda formula and gives a Chern number $-1$. The first term is related to the self-rotation of wave packet, and it equals to zero for closed-orbit subbands due to their parity symmetry. But, for open orbits, due to the divergence of saddle points, the singularity from the parent band needs to be taken into account.
Previously, Chang and Niu considered a simple square lattice model whose van Hove singularities (VHSs) are on the boundary of the Brillouin zone, and the number of VHSs is 2. Under a rational magnetic flux $p/q$, only one open-orbit subband emerges, and the first term gave Chern number $q$\cite{chang1996berry}.

For the case of $C\neq0$ Chern insulator, the VHSs of parent band become more complicate than the trivial case. As shown in FIG. \ref{fig4}(a), the VHSs shift from the boundary to the middle part of the Brillouin zone, and the number of VHSs becomes 4. Meanwhile, the electron pockets and hole pockets, separated by the green open orbits, become closer [FIG. \ref{fig4}(b)]. Under a rational magnetic flux $1/q$, the quantum tunnelling between the electron closed orbit and hole closed orbit gives birth to two open orbits with opposite propagating directions [FIG. \ref{fig4}(c,d)]. Consequently, these two open orbits carry Chern numbers $q-1$ and $-q-1$, respectively. We note that these two open orbits satisfy the general Diophantine equation but breaks the parameter regime of the TKNN consideration. This discrepancy indicates that the open orbits cannot adiabatically evolve into Landau levels.

\textit{Related to experiments.} --- The experiment advances in twist bilayer graphene\cite{andrei2020graphene,carr2017twistronics,finney2022unusual,yu2022correlated,saito2021hofstadter}, optical cold atom systems\cite{roati2008anderson,deissler2010delocalization,aidelsburger2013realization,schreiber2015observation,bordia2017probing,zhou2021thermalization} and hexagonal heterojunction structure\cite{dean2013hofstadter,hunt2013massive,ponomarenko2013cloning} provide a wide range of platforms to verify the Hofstadter spectrum of Chern insulator with exotic open orbits. 
On the other hand, the recent observation of quantum Hall effect in magic-angle twisted bilayer graphene have attracted extensive attention, in particular the anomalous Landau fan diagrams\cite{serlin2020intrinsic,stepanov2020untying,nuckolls2020strongly,wu2021chern,sharpe2019emergent,park2020flavour,stepanov2021competing}. Our finding about the Hofstadter butterfly of Chern insulator, including the relationship between the number of splitting subbands and Chern number of parent band and the appearance of anomalous open-orbit bands with Chern number $-q-1$, can bring new understanding on the mutual interaction between semiclassical quantization and topological properties of open orbits beyond the Landau level physics.

\section{Acknowledgements}
This work was financially supported by the National Basic Research Program of China (Grants No. 2017YFA0303301, No.2015CB921102), the National Natural Science Foundation of China (Grants No. 12022407, 11974256), and the Strategic Priority Research Program of the Chinese Academy of Sciences (Grant No. XDB28000000).
\bibliography{fly}

\end{document}